\documentstyle[11pt,psfig,epsf]{article}
\newcommand{\s}{\rm}

\newcommand{\ra}{\rightarrow}
\newcommand{\mn}{\mu \nu}

\newcommand{\be}{\begin{equation}}
\newcommand{\ee}{\end{equation}}

\newcommand{\bea}{\begin{eqnarray}}
\newcommand{\eea}{\end{eqnarray}}
\newcommand{\bef}{\begin{figure}}
\newcommand{\eef}{\end{figure}}

\begin{document}

\begin{flushright}
{\bf hep-ph/9910440}
\end{flushright}
\begin{center}
{\Large {Electromagnetic Signals of Hot Hadronic Matter}}
\footnote{Based on talk given by J.A. in the National Seminar on Nuclear
Physics, July 26-29, 1999, Institute of Physics, Bhubaneswar, India.}
\vskip 0.1in
Jan-e Alam$^{a,}$\footnote{On leave from Variable Energy Cyclotron Centre, 
1/AF Bidhan Nagar, Calcutta 700 064, India}, Sourav Sarkar$^b$, Pradip Roy$^c$, 
 T. Hatsuda$^a$\\ and Bikash Sinha$^{b,c}$\\
\vskip .1in

{\it a) Physics Department, Kyoto University, Kitashirakawa,
     Kyoto 606-8502, Japan}\\

{\it b) Variable Energy Cyclotron Centre,
     1/AF Bidhan Nagar, Calcutta 700 064
     India}\\

{\it c) Saha Institute of Nuclear Physics,
           1/AF Bidhan Nagar, Calcutta 700 064
           India}\\
\end{center}


\parindent=20pt
\vskip 0.2in
\begin{abstract}
The photon and dilepton emission rates from quark gluon plasma and
hot hadronic matter have been evaluated.  
The in-medium modifications of the particles 
appearing in the internal loop of the self energy diagram  
are taken into account by 
using a phenomenological effective Lagrangian approach, Brown-Rho
and Nambu scaling scenarios. 
We note that the in-medium effects on the low invariant mass 
distribution of dilepton and transverse momentum spectra of photon
are clearly visible.
\end{abstract}

\section*{I. Introduction}
Numerical simulations of the QCD (Quantum Chromodynamics) equation of state 
on the lattice predict that at very high density and/or temperature 
hadronic matter undergoes a phase transition to Quark Gluon Plasma 
(QGP)~\cite{ukawa}. One expects that ultrarelativistic heavy ion 
collisions (URHIC) might create conditions conducive for the 
formation and study of QGP. Various model calculations have been
performed to look for observable signatures of this state of matter.
However, among various signatures of QGP, photons and
dileptons are known to be advantageous,
primarily  so  because, electromagnetic interaction could
lead to detectable signal. However, it is weak enough to let the
produced particles (real photons and dileptons) escape the system
without further interaction and thus carrying the information 
of the constituents and their momentum distribution 
in the thermal bath.

The disadvantage with photons is the substantial background from 
various processes (thermal and non-thermal)~\cite{janepr}. Among these, 
the contribution from hard QCD processes is well understood in the 
framework of perturbative QCD and the yield from hadronic decays 
e. g. $\pi^0\,\ra\,\gamma\,\gamma$ can be accounted for by invariant mass 
analysis. However, photons from the thermalized hadronic gas pose a more
difficult task to disentangle. Therefore it is very important to estimate
photons from hot and dense hadronic gas along with the possible modifications
of the hadronic properties.   

We organize the paper as follows.
In section II we discuss the specific reactions which are used to calculate
the photon and dilepton emission rate  
from a thermal bath. In section III we discuss the medium 
modifications of hadrons. Section IV is devoted for evolution
dynamics. Results are presented in section V. 

\section*{II. Photon and Dilepton Emission Rates}
The basic aim of URHIC is to distinguish between the following two
possibilities:\\
\centerline{\bf{A\,+A\,$\ra$QGP$\ra$Mixed Phase$\ra$Hadronic Phase}}
\centerline{\,or\,}
\centerline{\bf{A\,+\,A\,$\ra$Hadronic Phase}}
The former (latter) case where the initial
state is formed in QGP (hadronic) phase will be called the `QGP scenario'
(`no phase transition scenario'). In the present work we will contrast
the photon and dilepton spectra originating from these two scenarios 
for SPS and RHIC energies.

The thermal emission rate 
of a real photon of energy $E$ and momentum $p$ 
can be expressed in terms of the trace of the retarded 
photon self energy ($\Pi_{\mn}^R$)at finite temperature~\cite{mt,gale}
\be
E\frac{dR}{d^3p}=-\frac{2}{(2\pi)^3}{\s {Im}}\Pi_\mu^{\mu\,R}\,(p)
\frac{1} {e^{E/T}-1}
\label{photrate1}
\ee
where $T$ is the temperature of the
thermal medium. The emission rate of thermal dileptons (virtual photon)
differs from that of real photon (due to different phase space factor)
in the following way,
\be
\frac{dR}{d^4q}=\frac{\alpha}{12\pi^4\,q^2}(1+\frac{2m^2}{q^2})
\sqrt{1-\frac{4m^2}{q^2}}{\s Im}\Pi^{R\mu}_\mu\,
\frac{1} {e^{q_0/T}-1}
\label{drd4q2}
\ee

The photon emission rate due to Compton and annihilation processes
in QGP was performed  
in Refs.~\cite{kapusta,baier} by applying the Hard Thermal
Loop (HTL) resummation \cite{bp,ft}.
The rate of hard photon ($E>T$) emission  
due to these processes is given by~\cite{kapusta}
\be
E\frac{dR_\gamma^{QGP}}{d^3q}=\frac{5}{9}\frac{\alpha\alpha_s}{2\pi^2}
T^2\,e^{-E/T}\ln(2.912E/g^2T).
\label{p1}
\ee

Recently, the bremsstrahlung contribution to photon emission 
rate has been computed~\cite{aurenche} 
by evaluating the photon self energy in two loop HTL approximation.
The physical processes arising from two loop 
contribution are the bremsstrahlung of quarks, antiquarks and
quark anti-quark annihilation with scattering in the thermal bath. 
The rate of 
photon production due to bremsstrahlung process for a two flavor
thermal system with $E>T$ is given by~\cite{aurenche}
\be
E\frac{dR_\gamma^{QGP}}{d^3q}=\frac{40}{9\pi^5}\,\alpha\alpha_s
T^2\,e^{-E/T}\left(J_T-J_L\right)\ln 2,
\label{p2}
\ee
and the rate due to $q-\bar q$ annihilation with scattering 
in the thermal bath is given by,
\be
E\frac{dR_\gamma^{QGP}}{d^3q}=\frac{40}{27\pi^5}\,\alpha\alpha_s
ET\,e^{-E/T}\left(J_T-J_L\right),
\label{p3}
\ee
where $J_T\approx 4.45$ and $J_L\approx -4.26$.
The most important implication of this work is that the two loop
contribution is of the same order of magnitude as those evaluated
at one loop~\cite{kapusta,baier} due to the larger size of the
available phase space. The net rate of emission is obtained
by adding eqs. \ref{p1}, \ref{p2} and \ref{p3}.

In the hadronic matter (HM) an exhaustive set of hadronic reactions 
and vector meson decays involving $\pi$, $\rho$, $\omega$ and
$\eta$ mesons have been considered. 
It is well known~\cite{kapusta} 
that the reactions $\pi\,\rho\,\ra\, \pi\,\gamma$ , 
$\pi\,\pi\,\ra\, \rho\,\gamma$ , $\pi\,\pi\,\ra\, \eta\,\gamma$ , 
$\pi\,\eta\,\ra\, \pi\,\gamma$ , and the decays $\rho\,\ra\,\pi\,\pi\,\gamma$
and $\omega\,\ra\,\pi\,\gamma$ are the most important channels 
for photon production from hadronic matter in the
energy regime of our interest. The rates for these processes could be 
evaluated from the imaginary part of the two loop photon self energy
involving various mesons. 
The photon emission rate from $\pi\,\rho\,\ra\, \pi\,\gamma$ via
the intermediary $a_1$ has also been included.
 
We have considered quark anti-quark annihilation 
for the evaluation of dilepton emission rate from QGP,
which is given by
\be
\frac{dR}{dM} = \frac{\sigma_{q\bar q}(M)}{(2\pi)^4}\,M^4\,T\,\sum_n
\,\frac{K_1(nM/T)}{n}
\ee
with the  cross section 
\be
\sigma_{q\bar{q}\rightarrow e^+e^-}=\frac{80\pi}{9}\frac{\alpha^2}{M^2}
\sqrt{\left(1-\frac{4m^2}{M^2}\right)}
\,\left(1+\frac{2m^2}{M^2}\right).
\ee
For low mass dilepton emission rate we consider the yield from 
$\pi^+\pi^-\,\ra\,e^+e^-$, 
$\rho\,\ra\,e^+e^-$ and $\omega\,\ra\,e^+e^-$ (see ~\cite{ja,pr} for
details). 

\section*{ III. Medium Effects}
To study the medium effects on the transverse momentum distribution 
of photons and invariant mass distribution of dileptons from 
URHIC we need two more ingredients. Firstly, we require
the variation of masses and decay widths with temperature, because
the invariant matrix element for photon (real and virtual)
production suffer in-medium
modifications through the temperature dependent masses and
widths of the participants. 
As the hadronic masses and decay widths enter directly in the 
count rates of electromagnetically interacting particles, the finite
temperature effects in the cross sections, particularly
in the hadronic matter are very important in URHIC.

We consider the following scenarios  
for the in-medium vector meson mass variations in the present work.

\noindent{{\bf (I) Walecka Model}\,\cite{vol16}\\
We have evaluated the in-medium mass of $\rho$ and $\omega$
mesons and decay widths within the framework of Walecka model.
The renormalization procedure of Refs.~\cite{sh,hsk} has
been used to render the vacuum self energy of vector mesons
finite. The details of the  calculations at non-zero $T$ can be found in our
previous works~\cite{ja,pr,sourav,npa99,jsphs} 
and we do not reproduce them here.
However, from the results of those calculations 
the variation of nucleon, rho and omega masses and the 
decay width($\Gamma_\rho$)
of rho with temperature at zero baryon density can be 
parametrized as (see also~\cite{prc})
\be
m_N^\ast/m_N=1-0.0264x^{8.94},\nonumber
\hskip 0.4cm
m_\rho^\ast/m_\rho=1-0.1268x^{5.24}\nonumber
\ee
\be
m_\omega^\ast/m_\omega=1-0.0438x^{7.09},\nonumber
\hskip 0.4cm
\Gamma_\rho^\ast/\Gamma_\rho=1+0.664x^4-0.625x^5
\ee
where asterisk indicates effective mass/width in the medium,
$x=T/T_c$ and $T_c=0.16$ GeV. 
According to our calculation the nucleon, rho and omega masses decrease
differently. 

\noindent{{\bf (II) Brown-Rho and Nambu Scaling}~\cite{brprl,brpr}}\\
In the previous it is mentioned that both the photon and dilepton
emission rates are proportional to imaginary part of the 
current-current correlation functions (spectral function).
We have parameterized the hadronic spectral functions
with a conspicuous resonance plus a continuum. These spectral
functions (in vacuum) for the isovector and isoscalar channel have been
constrained from experimental data on $e^+e^-\rightarrow hadrons$.
At finite temperature we have parameterized the vector meson
mass (pole) and the continuum threshold ($\omega_0$) of the spectral function 
as a function of temperature
in the following way (see Ref.~\cite{jsphs} for details):
\be
{m_{V}^* \over m_{V}}  = 
{\omega_{0}^* \over \omega_{0}}  =
 \left( 1 - {T^2 \over T_c^2} \right) ^{\lambda},
\label{anst}
\ee
where $\lambda$ is a sort of {\em dynamical} critical exponent and
$V$ stands for vector mesons ($\rho$ and $\omega$).
 Since the numerical value of $\lambda$ is not known, 
 we take two typical cases:
 $\lambda=1/6$ (BR scaling) and $1/2$ 
 (Nambu scaling)~\cite{brpr}.

\section*{IV. Evolution Dynamics}

The observed photon spectrum originating from an expanding 
QGP or hadronic matter is obtained by convoluting the static
(fixed temperature) rate, as given by Eqs.~(\ref{photrate1})
and ~(\ref{drd4q2}), with expansion dynamics. 
Therefore, the second ingredient required for our calculations is 
the description
of the system undergoing rapid expansion from its initial formation
stage to the final freeze-out stage.
In this work we use Bjorken-like~\cite{bjorken}
hydrodynamical model for the isentropic expansion of the matter
in ($1 + 1$) dimension.
For the QGP sector we use a simple bag model equation of state (EOS) with
two flavor degrees of freedom. The temperature in the QGP phase evolves
according to Bjorken scaling law $T^3\,\tau=T_i^3\tau_i$.
In the hadronic phase we have to be more careful about the presence
of heavier particles and their change in masses due to finite temperature
effects.
The hadronic phase consists of $\pi$, $\rho$, $\omega$, $\eta$ and $a_1$ 
mesons and nucleons. The nucleons and heavier mesons may play an important
role in the EOS in a scenario where mass of the hadrons decreases
with temperature. 
The energy density and pressure
for such a system of mesons and nucleons is given by,
\be
\epsilon_H=\sum_{i=mesons} \frac{g_i}{(2\pi)^3} 
\int d^3p\,E_i\,f_{BE}(E_i,T)
+\frac{g_N}{(2\pi)^3} 
\int d^3p\,E_N\,f_{FD}(E_N,T)
\ee
and
\be
P_H=\sum_{i=mesons} \frac{g_i}{(2\pi)^3} 
\int d^3p\frac{p^2}{3\,E_i}f_{BE}(E_i,T)
+\frac{g_N}{(2\pi)^3} 
\int d^3p\frac{p^2}{3\,E_N}f_{FD}(E_N,T)
\ee
where the sum is over all the mesons under consideration and $N$ stands
for nucleons and $E_i=\sqrt{p^2 + m_i^2}$.          
The entropy density is given by
\be
s_H=\frac{\epsilon_H+P_H}{T}\,\equiv\,4a_{\s{eff}}(T)\,T^3
= 4\frac{\pi^2}{90} g_{\s{eff}}(m^\ast(T),T)T^3
\label{entro}
\ee
where  $g_{\s{eff}}$ is the effective statistical degeneracy.

Thus, we can visualize the effect of finite mass of the hadrons
through an effective degeneracy $g_{\s{eff}}(m^\ast(T),T)$ of the 
hadronic gas. The variation 
of temperature from its initial value  $T_i$ to final value 
$T_f$ (freeze-out temperature) with proper time ($\tau$) is governed 
by the conservation of entropy 
\be
s\tau=s_i\tau_i 
\label{entro1}
\ee
The initial temperature of the system is obtained by solving
the following equation self consistently
\be
\frac{dN_\pi}{dy}=\frac{45\zeta(3)}{2\pi^4}\pi\,R_A^2 4a_{\s{eff}}T_i^3\tau_i
\label{dnpidy}
\ee
where $dN_\pi/dy$ is the total pion multiplicity, $R_A$ is the radius
of the system, $\tau_i$ is the initial thermalization time and 
$a_{\s{eff}}=({\pi^2}/{90})\,g_{\s{eff}}$. 
The change in the expansion dynamics
as well as the value of the initial temperature due
to medium effects enters the calculation of the
photon and dilepton emission rates through the effective statistical degeneracy.

\section*{V. Results and Discussions}
Having obtained the finite temperature effects on hadronic properties
and the cooling laws we now
integrate the rates obtained in the previous sections over the space-time
evolution of the collision. We must account for the fact that the thermal rates
are evaluated in the rest frame of the emitting matter and hence the momenta of the
emitted photons or dileptons are expressed in that frame. Accordingly,
the integral over the expanding matter is of the form
\be
\frac{dN}{d\Gamma}=\begin{array}{c}
{\small {freeze-out}}\\{\displaystyle\int}\\{\small {formation}}\end{array}\,
d^4x\frac{dR(E^\ast,T(x))}{d\Gamma}
\ee 
where $d\Gamma$ stands for invariant phase space elements:
$d^3p/E$ for photons and $d^4q$ for dileptons.
$E^\ast$ is the energy of the photon or lepton pair 
in the rest frame of the emitting matter
and $T(x)$ is the local temperature.
In a fixed frame like the laboratory or the centre of mass frame, where
the 4-momentum of the photon or lepton pair is $q_\mu=(E,\vec q)$ 
and the emitting matter element $d^3x$ moves with a velocity
$u_\mu=\gamma(1,\vec v)$, the energy in the rest frame of the fluid 
element is given by $E^\ast=u_\mu q^\mu$.
\begin{figure}
\centerline{\psfig{figure=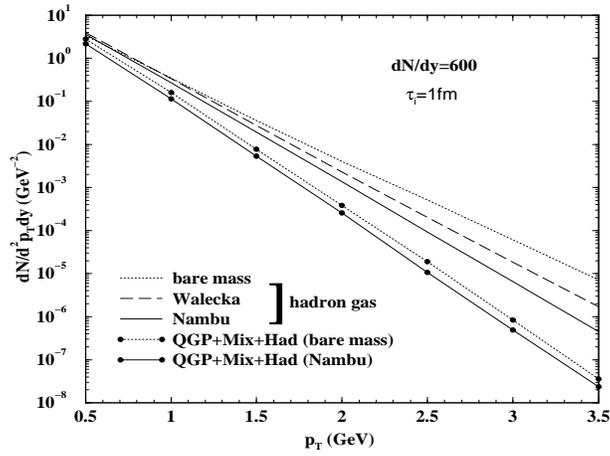,height=6cm,width=8cm}}
\caption{Total thermal photon yield corresponding to $dN/dy=600$ and
$\tau_i=1$ fm/c. 
The solid (long-dash) line indicates photon spectra  
when hadronic matter formed in the initial state 
at  $T_i=195$ MeV ($T_i=220$ MeV) 
and the medium effects are taken from Nambu scaling (Walecka model).
The dotted line  represents the photon spectra without 
medium effects
with $T_i=270$ MeV. The solid (dotted) line with solid dots
represent the yield for the `QGP scenario' when the hadronic
mass variations are taken from Nambu scaling (free mass).
}
\label{7fig17}
\end{figure}

As discussed earlier, $g_{\s eff}$ is obtained as a function of $T$  
by solving Eq.(\ref{entro}). A smaller (larger) value of $g_{\s eff}$ 
is obtained in the free (effective) mass scenario. 
As a result we get a larger (smaller) initial temperature
by solving Eq.(\ref{dnpidy}) in the free (dropping) mass scenario 
for a given multiplicity.
Naively we expect that at a given temperature if a meson mass drops
its Boltzmann factor will be enhanced and more of those mesons will
be produced leading to more photons~\cite{sourav}.
However, a larger drop in the hadronic masses results in smaller initial
temperature, implying that the
space time integrated spectra crucially depends on these two
competitive factors.
Therefore, with (without) medium effects one integrates an enhanced (depleted) 
static rate over smaller (larger) temperature range for a fixed
freeze-out temperature ($T_f=130$ MeV in the present case).
In the present calculation 
(Fig.~(\ref{7fig17}) for Pb+Pb collisions at SPS energies, $dN/dy=600$) 
the enhancement in the photon emission due to 
the higher initial temperature in the free mass scenario (where static rate is
smaller) overwhelms the enhancement of the rate due to negative shift
in the vector meson masses (where the initial temperature is smaller).
Accordingly, in the case of bare mass (Nambu scaling) scenario the 
photon yield is the highest (lowest). In case of the Walecka model,
the photon yield lies between the above two limits.
\begin{figure}
\centerline{\psfig{figure=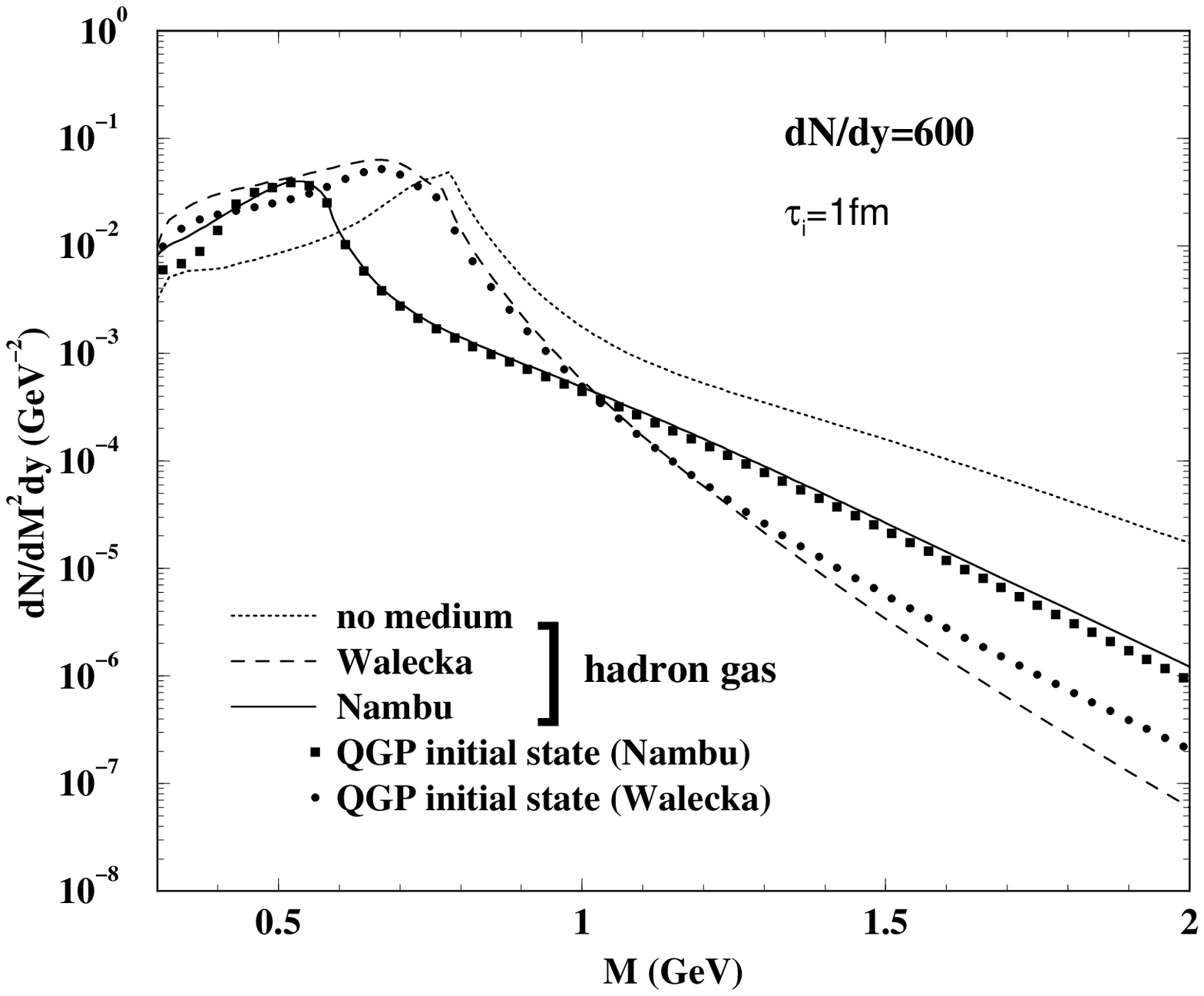,height=6cm,width=8cm}}
\caption{Total thermal dilepton yield corresponding to $dN/dy=600$ and
$\tau_i=1$ fm/c. 
The solid (long-dash) line indicates dilepton spectra  
when hadronic matter formed in the initial state 
at  $T_i=195$ MeV ($T_i=220$ MeV) 
and the medium effects are taken from Nambu scaling (Walecka model).
The dotted line  represents the spectra without medium effects with
$T_i=270$ MeV.
The square (solid dots) line with solid dots
represent the yield for the `QGP scenario' when the hadronic
mass variations are taken from Nambu scaling (Walecka model).
}
\label{7fig18}
\end{figure}

In the `QGP scenario' the  photon
yield with in-medium mass is lower than the case 
where bare masses of hadrons are
considered. However, the difference is considerably less than the
`no phase transition scenario'.   
This is because, in this case
the initial temperature is determined by the quark and gluon 
degrees of freedom and the only difference between the two
is due to the different lifetimes of the mixed phase.  
In Fig.~(\ref{7fig17}), the photon spectra
from `QGP scenario' is compared with that from `no phase
transition scenario'; the latter overshines the former.

The space time integrated dilepton spectra  for
the `QGP scenario' and `no phase transition scenario' 
with different mass variation are shown in Fig.(\ref{7fig18}).
The shifts in the invariant mass distribution of the spectra
due to the reduction in the hadronic masses according
to different models are distinctly visible. Similar to the
photon spectra, the dilepton spectra from `no phase
transition scenario' dominates over the `QGP' scenario
for invariant mass beyond $\rho$ peak.

\begin{figure}
\centerline{\psfig{figure=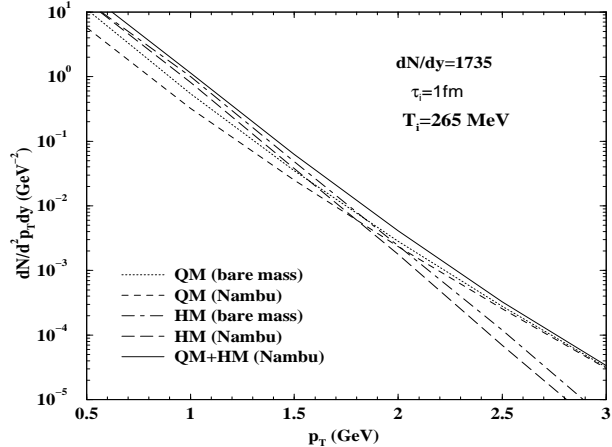,height=6cm,width=8cm}}
\caption{Thermal photon spectra at RHIC energies.
}
\label{7fig20}
\end{figure}

For RHIC energies, we  consider the `QGP' scenario only as 
a scenario of pure hot hadronic initial state
within the format of the model used here appears to
be unrealistic. 
The thermal photon yield for RHIC ($dN/dy=1735$) 
is displayed in Fig.~(\ref{7fig20}). 
The solid line represents the total thermal photon yield 
originating from initial QGP state, mixed phase and the pure hadronic
phase. The short dash line indicates photons from quark 
matter (QM) (= pure QGP phase + QGP part of the mixed phase)
and the long dash line represents photons from hadronic matter (HM)
(= hadronic part of the mixed phase + pure hadronic phase).
In all these cases the effective masses of the hadrons have been taken 
from Nambu scaling. For $p_T>2$ GeV photons from QM overshines
those from HM, since most of these high $p_T$ photons originate
from the high temperature QGP phase.
The dotted and the dotdash lines indicate photon yields from QM
and HM respectively with bare masses in the hadronic sector. 
The HM contribution for the bare mass is larger than the
effective mass (Nambu) scenario because of the larger value of the life
time of the mixed phase in the earlier case.
It is important to note that for $p_T>2$ GeV, the difference in the QM 
and HM contribution in the effective mass scenario is more than the
bare mass scenario.

Thermal dilepton yield at RHIC energies for QGP initial state 
and for different mass variation scenarios are shown in 
Fig.~(\ref{7fig21}). The shape of the peak in the dilepton
spectra in case of Walecka model is slightly different(broader) from the other
cases because of the larger mass separation between $\rho$ and $\omega$
mesons in this case. The dilepton yield beyond the vector meson
peak is larger in the bare mass scenario because of the larger 
life time of the mixed phase.
\begin{figure}
\centerline{\psfig{figure=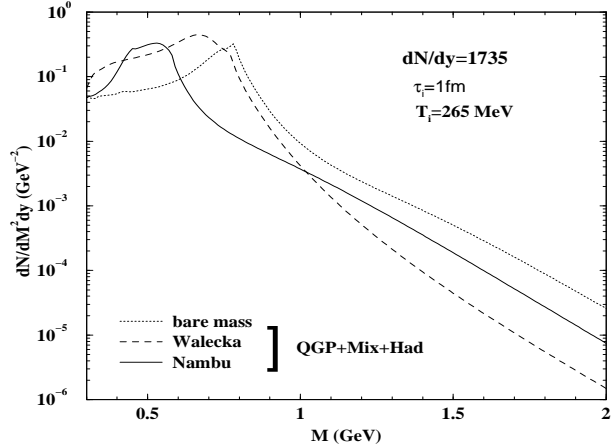,height=6cm,width=8cm}}
\caption{Thermal dilepton spectra at RHIC energies.
}
\label{7fig21}
\end{figure}

We have evaluated both the real photon and dilepton emission rate
from QGP and hot hadronic matter by taking into accounts in-medium
mass modifications of vector mesons according to Walecka Model
calculations, BR and Nambu scaling. We observe that the in-medium
effects on both photon and dilepton spectra are clearly visible. 

{\bf Acknowledgement:} J. A. is grateful to Japan Society
for Promotion of Science (JSPS) for financial support.
T. H. was partly supported by Grant-in-Aid for Scientific
Research No. 10874042 of the Japanese Ministry of Education,
Science, and Culture. J. A. and T. H. were also supported 
 by Grant-in-Aid for Scientific Research No. 98360 of
 JSPS.

\end{document}